# Diffusion-induced anisotropic cancer invasion: A novel experimental method based on tumor spheroids

Rosalia Ferraro[1,2] | Flora Ascione[1] | Prashant Dogra[3,4] | Vittorio Cristini[3,5,6] | Stefano Guido[1,2] | Sergio Caserta[1,2]

[1]Università degli Studi di Napoli Federico II, Dipartimento di Ingegneria Chimica, dei Materiali e della Produzione Industriale, Naples, Italy

[2]CEINGE Advanced Biotechnologies, Naples, Italy

[3]Mathematics in Medicine Program, Houston Methodist Research Institute, Houston, Texas, USA

[4]Department of Physiology and Biophysics, Weill Cornell Medical College, New York, New York, USA

[5]Department of Imaging Physics, University of Texas MD Anderson Cancer Center, Houston, Texas, USA

[6]Physiology, Biophysics, and Systems Biology Program, Graduate School of Medical Sciences, Weill Cornell Medicine, New York, New York, USA

**Correspondence**
Sergio Caserta: Università degli Studi di Napoli Federico II, Dipartimento di Ingegneria Chimica, dei Materiali e della Produzione Industriale, Piazzale Vincenzo Tecchio, 80, 80125 Napoli NA, Italy.
Email: sergio.caserta@unina.it

**Present address**
Flora Ascione, IFOM, FIRC Institute of Molecular Oncology, Milan, Italy

**Abstract**

Tumor invasion is strongly influenced by microenvironment and, among other parameters, chemical stimuli play an important role. An innovative methodology for the quantitative investigation of chemotaxis *in vitro* by live imaging of morphology of cell spheroids, in 3D collagen gel, is presented here. The assay was performed by using a chemotactic chamber to impose a controlled gradients of nutrients (glucose) on spheroids, mimicking the chemotactic stimuli naturally occurring in the proximity of blood vessels. Different tumoral cell lines (PANC-1 and HT-1080) are compared to non-tumoral ones (NIH/3T3). Morphology response is observed by means a Time-lapse workstation equipped with an incubating system and quantified by image analysis techniques. Description of invasion phenomena was based on an engineering approach, based on transport phenomena concepts. As expected, NIH/3T3 spheroids are characterized by a limited tendency of cells to invade the surrounding tissue, unlike PANC-1 and HT-1080 that show relatively stronger response to gradients.

**KEYWORDS**

chemotaxis, diffusional instability, time-lapse microscopy, transport phenomena, tumor spheroids

## 1 | INTRODUCTION

In a tumor, diseased cells proliferate uncontrollably, compared to healthy cells, and disrupt tissue organization. Sometimes, cancer cells can migrate from the main tumor body, invade and grow at a distant site by a process known as metastasis. Such expansion of the tumor, which was initially localized in a specific site of the body, heavily reduces patient survival. In particular, cancer cell invasion can be induced by nutrients gradients diffusing from blood vessels to surrounding tissues, or inverse catabolite gradients from tumor core,[1,2] in accordance with a process described in the literature by the diffusion instability model.[3] According to this model, instability can allow tumors to grow indefinitely: through heterogeneous cell proliferation and migration, microenvironmental substrate gradients, for example, of cell nutrients or the extracellular matrix, may drive tumor invasion through morphological instability with separation of cell clusters from the tumor edge and infiltration into surrounding normal tissue.[4]

The diffusion instability model is based on deterministic analysis of clinical data,[5–8] interpreted according to physical principles.[3,4,9–12] It can predict growth, invasion, and morphologic instability of tumors, describing the process according to biophysical laws, regulated by heterogeneity in phenotypic, genotypic, and microenvironmental parameters.[4,13] Quantitative evolution of tumor mass is described by a set of differential equations,[3] where the density of each cell type (healthy or cancer) co-populating a tissue is described by a mass balance where cell density can increase or decrease depending upon motility, chemotaxis, growth, death, necrosis, and mutation.[14] Each of these phenomena depends on nutrient availability and catabolite concentration that is defined by flows from/to blood vessels, and consumption/generation rates due to cell metabolism. Competition for nutrients among different cell types, with cancerous cells being more adaptive, resilient, proliferative, and metabolically active, with respect to healthy cells, forms the basis of this model.[3,15]

The process by which cells move in response to external chemical concentration gradients is known as chemotaxis.[16–20] The study of chemotaxis *in vitro* raised growing interest in the literature, but still deserve further systematic investigations. Several experimental approaches have been proposed over years to investigate, both qualitatively and quantitatively, single cell chemotaxis,[18,20–22] for example, Boyden chamber,[23] agarose assay,[21] and microfluidic devices, usually fabricated in polydimethylsiloxane (PDMS).[24–28]

An important topic in cancer research is study of chemotaxis in a quantitative way using 3D tumor models capable of replicating *in vitro* the tumor microenvironment (TME) found *in vivo*. In fact, 3D models retain the primary physiological features of human cancer[29] (e.g., key cell–cell interactions, extracellular matrix [ECM] deposition by cells occurring *in vivo*, concentration gradients, and necrosis), which are lacking in simpler 2D models.[27,30,31] One of the most used 3D models to investigate cancer growth is based on tumor spheroids,[30,32,33] tightly bound cellular aggregates able to mimic non-vascularized cancer mass,[32–34] typical of initial stages of the pathology. Tumor spheroid models, able to reproduce conditions typical of the *in vivo* environment, are suitable for development of *ex vivo* models.

In this paper, a novel *in vitro* tumor invasion assay, based on tumor spheroids seeded in a 3D collagen gel to mimic the ECM,[20] is proposed. The entire scaffold is hosted in a microfluidic chemotactic chamber designed to generate external chemical stimuli, such as concentration gradients, present in real *in vivo* conditions.[19,20] Spheroids dynamics evolution is monitored by means of a live cell imaging workstation, time-lapse microscopy (TLM). The novel approach here presented investigates the role of transport phenomena, in particular diffusive flows, and related chemical stimuli, on cancer invasion. In detail, invasiveness is quantified by measuring different morphological (area and deformation) and mechanical (motility and invading cells) parameters and by relating them to nutrient (glucose) gradient to which the spheroids are subjected. Consequently, the relationship between the chemoattractant gradient and mechanical response of cells is established. This relationship is known as mechanochemistry, which investigates mechano-sensing and signaling pathways of chemical cues that activate the tumor mechanical behaviors.[35–40] In order to quantify the correlation, controlled diffusive flow of glucose was calculated using a finite element *in silico* model[3,15,19,41] of the experimental system.

As it is known, glucose is one of the components of cell culture medium and stimulate cell metabolism[42]; its role in promoting cancer invasion has been previously observed.[43–46] Therefore, glucose is expected to induce the invasion of cancer cells from spheroids into the surrounding environment. As a control, uniform concentration of fetal bovine serum (FBS) is used to mimic standard cell growth conditions (FBS 10%) and study dynamic behavior of spheroids and potential invasiveness of tumor cells in the surrounding: molecules and growth factors contained in FBS can stimulate single cell growth and migration.[19] In addition, the dynamic behavior of cell spheroids in a nutrient-depleted environment (FBS 0% and low glucose concentration) is also investigated as negative control. All the experiments are performed on two cancer cell lines: HT-1080 human fibrosarcoma cells and PANC-1 human pancreatic carcinoma cell. A non-tumor cell line (NIH/3T3 mouse embryonic fibroblasts) is used as control.

Invasion assay proposed in this paper provides an innovative tool to investigate the role of transport phenomena, and in particular diffusive mechanisms, on dynamic evolution, growth and invasiveness of cells system, with specific focus on cancer invasion. In other terms, to the best of our knowledge, the methodology here proposed represent the first clear and systematic application *in vitro* of the diffusional instability model,[3,4,6,15] ever reported in the literature. The knowledge of these basic mechanisms here investigated, and the methodology here presented can be used to develop *ex vivo* experimental models to investigate the response of a tumor to precision therapeutic treatments, with potential implementation in personalized medicine.

## 2 | MATERIALS AND METHODS

### 2.1 | Cell culture

Two cancer cell lines (PANC-1 and HT-1080) and a non-tumor one (NIH/3T3) were used in the experiments. All cells were cultured in their standard growth medium in 2D monolayers under the typical cell culture conditions, at 37°C in a humidified atmosphere (5% $CO_2$).

NIH/3T3 mouse fibroblasts and PANC-1 human pancreatic carcinoma cells were cultured in Dulbecco's Modified Eagle's Medium (DMEM) supplemented with 10% (v/v) FBS, 1% (v/v) antibiotics (50 units/mL penicillin and 50 mg/mL streptomycin), and 1% (v/v) L-glutamine. HT-1080 human fibrosarcoma cells were cultured in Eagle's Minimum Essential Medium (EMEM) supplemented with 10% (v/v) FBS, 1% (v/v) antibiotics, and 1% (v/v) L-glutamine.

### 2.2 | Spheroids formation

Spheroids were produced using the "agarose multi-well plate method".[26,29,47] Specifically, 1% agarose solution was prepared by dissolving agarose powder (E AGAROSE, Conda, Cat n° 8100) in water at 200°C for 20 minutes using a magnetic stirrer to homogenize the solution. Then, agarose solution was rapidly pipetted in 200 μL aliquots into the wells of a 48-well culture dish (Nunclon Δ Multidishes, 48 wells, flat bottom, Thermo Scientific, Nunc 150687) under sterile conditions and allowed to cool down. By capillary, agarose solution rises along the walls of the wells, thus gelifying in a few minutes forming a hemispherical meniscus. The non-adhesive concave surface promotes the collection of cells in the meniscus and cell–cell adhesion establishment; this results in the formation of cell aggregates and finally spheroids after an adequate incubation time, depending on cell type and concentration.

To generate spheroids, cells were harvested from monolayer cultures and counted. $2 \cdot 10^3$ cells were seeded in single wells pre-coated with non-adhesive agarose and covered with the specific cell growth medium (see Section 2.1). The multi-well plate was then incubated under typical cell culture conditions at 37°C to allow spheroid formation. Typically, 5–10 days (depending on cell line) were required to obtain compact spheroids of adequate size.

## 2.3 | Time-lapse microscopy

Tumor invasion process was monitored by using an automated TLM workstation,[14,48] based on an inverted microscope (Zeiss Axiovert 200, Carl Zeiss, Jena, Germany) enclosed in a homemade incubator, capable to guarantee physiological conditions to ensure cell viability: constant temperature (37 ± 0.1°C), 5% $CO_2$ and 100% humidified atmosphere. The workstation was also equipped with motorized stage and focus control (Microscope Stages - Märzhäuser Wetzlar) for automated positioning, allowing to iteratively image specific regions, and was controlled by homemade control software in LabVIEW (National Instruments). Images were acquired by a CCD video camera (Orca AG, Hamamatsu, Japan) using a long working distance 5× objective in phase contrast (EC Plan - NEOFLUAR Ph1), at regular intervals (20 min) for 48 h.

## 2.4 | Chemotaxis chamber

Chemotaxis chambers used in this work were described and characterized in previous works.[19,20] Two chemotaxis chambers, each hosting two different samples in two separate wells, were used at the same time to run four tests in parallel during each experiment. In each well, a porous membrane (0.22 μm pores, MF-Millipore), sandwiched between two rectangular metal plates (Figure 1A), separated two sections of the chamber. One section was loaded with a spheroids-seeded collagen gel (sample well), and the other was used as a reservoir of the chemoattractant solution (chemoattractant well; Figure 1B). The 2-mm hole in the metal plates realizes a cylindrical connection between the two chambers, hosting the membrane. The wells were sealed on the bottom by gluing under each of the two wells a glass slide (24 mm × 24 mm), using a mixture of vaseline and paraffin (ratio 1:1 in weight).

## 2.5 | Experimental design

FBS and glucose were used in the experiments as nutrient sources. Three experimental conditions were explored and compared to quantitatively investigate the effect of FBS and glucose gradients on tumor invasion:

- ISO 10%: Isotropic condition with 10% FBS in order to mimic standard cell culture conditions. In this experimental condition, the membrane sandwiched between the two metal plates was not added, and the entire chamber was loaded with collagen solution containing 10% FBS and embedded with cell spheroids.
- ISO 0%: Isotropic condition without FBS (0%) to evaluate the morphological response of cell spheroids under isotropic stress conditions. In this experimental condition, the sample well of the chemotaxis device was loaded with a spheroids-seeded collagen gel, containing no FBS and a low concentration of glucose (see Section 2.6). The chemoattractant reservoir contained FBS-free medium diluted in water at the same concentration (v/v) of FBS-free medium in collagen gel solution in order to keep uniform and low the concentration of glucose (contained in standard medium) during the experiment. This experimental condition can be considered as a *negative* control.
- $\nabla C$: Invasion assay, in which cell spheroids were exposed to a controlled diffusive flow of glucose. In this experimental setup, the sample well of the chemotaxis device was loaded with a spheroids-seeded collagen gel, containing no FBS and a low concentration of glucose (see Section 2.6). The chemoattractant reservoir was filled with a solution of glucose 10 g/L. In this condition, cell spheroids were under controlled diffusional stress, in an environment mimicking physiological conditions where nutrients diffusive flows had a preferential directional orientation.

## 2.6 | Preparation of collagen gel

Cell spheroids were suspended in collagen gel solution under sterile conditions. In ISO 10% condition collagen gel was prepared by mixing collagen (final concentration 2 mg/mL), 10× phosphate buffered saline (PBS) at a final concentration of 10% in volume, FBS-free DMEM, and FBS at a final concentration of 10% in volume.

In ISO 0% and $\nabla C$ conditions, a low glucose concentration was imposed in order to mimic a nutrient-depleted environment. Specifically, collagen gel was prepared by mixing collagen (final concentration 2 mg/mL), 10× PBS at a final concentration of 10% in volume, FBS-free DMEM (10% in medium, see Section 2.1), diluted 1:1 with deionized water to further reduce the glucose concentration. In this case, the concentration of glucose in collagen gel resulted to be 1.5 g/L, being 4.5 g/L the concentration of glucose in DMEM.

In all conditions, 1 M NaOH was added to the solution to adjust the pH to a value between 7 and 7.4.

## 2.7 | Preparation of assays

For $\nabla C$ and ISO 0% conditions, 180 μL of collagen solution were poured in the sample well of the chemotaxis chamber, previously sterilized in autoclave, and let to partially gelify for 20 min at room temperature. Instead, for ISO 10% condition, 350 μL of collagen solution was used. The greater amount of collagen solution in the ISO 10% condition was needed to pour out the higher volume of the chamber, due to the absence of the two metal plates and membrane. On the contrary, the reservoir, obtained by inserting the membrane sandwiched between the two metal plates, was filled with the collagen solution for the ISO 0% condition and with the chemoattractant for the ∇C condition, which generates a concentration gradient in the sample well. The chemoattractant was obtained by mixing 10 g/L of glucose in 10 mL of a solution containing water and FBS-free medium (2:1).

Cell spheroids grown on agarose substrate (see Section 2.2) were picked by using a pipette and suspended in collagen gel solution. For ∇C and ISO 0% conditions, 200 μL of collagen solution containing a significant number of cell spheroids (∼20) were poured on the lower spheroids-free collagen gel and let polymerize for 40 min at room temperature. For ISO 10% condition, 400 μL of collagen solution containing cell spheroids were used. Spheroids slowly sedimented randomly along the chamber, while collagen gelification protocol completed. After gelification, standard culture medium at the proper composition (depending on the experimental condition) was added to fill the sample well to prevent gel dehydration.

In Figure 1C, a top view scheme of the chemotaxis chamber[19,20] is reported. Dark square on the top represents the chemoattractant

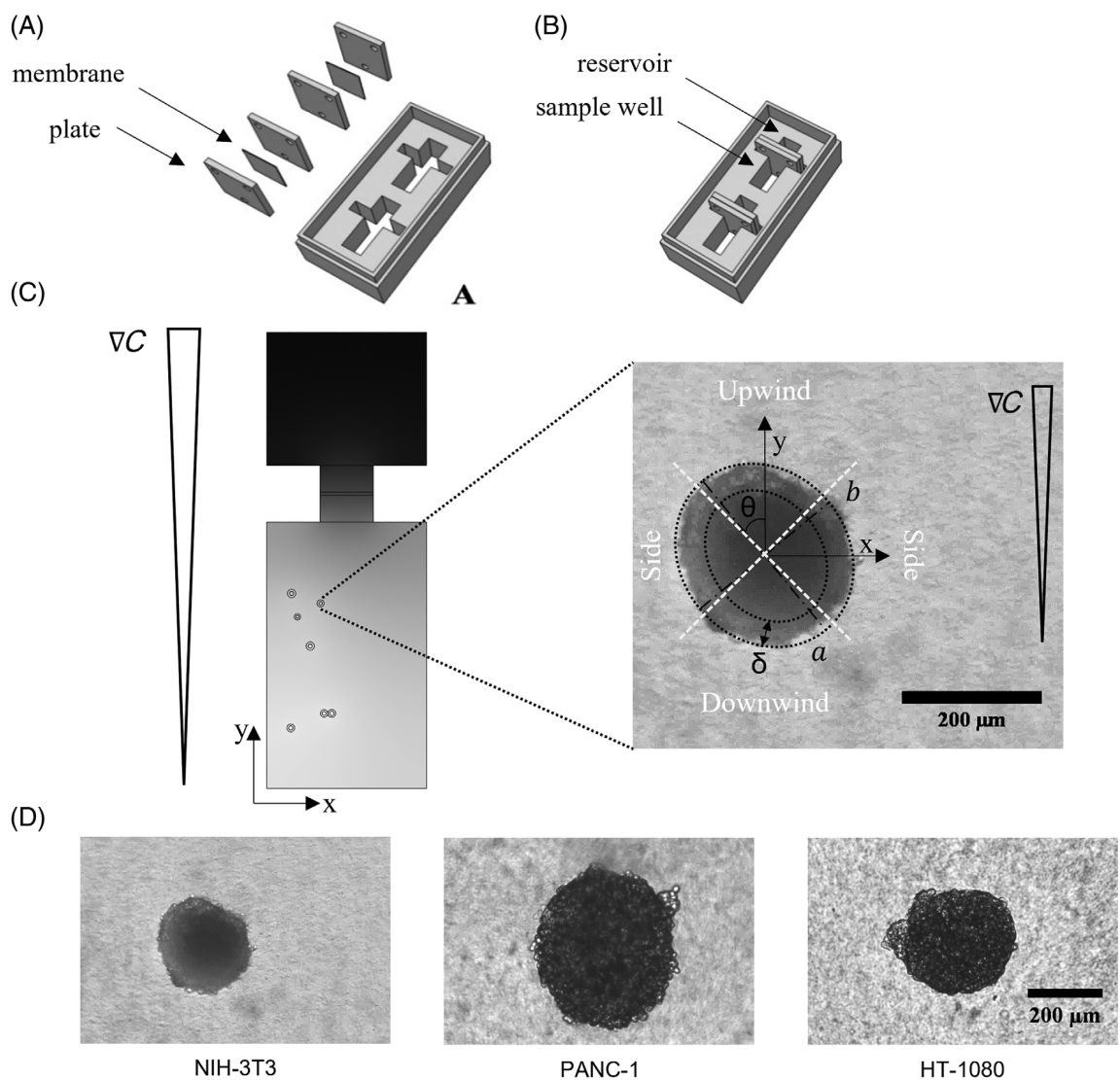

**FIGURE 1**  (A) In the exploded view rendering of the chamber, all the components are individually visible, in particular, the porous membrane and the two rectangular metal plates. (B) In the assembled rendering, the membrane separates the sample well and the chemoattractant reservoir. (C) Schematic of the chemotaxis chamber used for numerical simulation includes two main areas: the chemoattractant well (dark rectangle on the top) and spheroids-seeded collagen gel section (bottom). The dynamic of the tumor invasion assay was quantified by measuring different morphological parameters as shown in the image (C, right): (i) spheroid and necrotic core area (outer and inner dotted black circles, respectively), (ii) major and minor axis ($a$ and $b$, dashed dotted black lines), (iii) orientation angle ($\theta°$) between the major axis and the vertical direction, that is, parallel to the diffusive flow, and (iv) viable rim thickness ($\delta$). In the image are also reported in white the three sections defined as upwind, downwind, and side (demarcated by dotted white lines). (D) Representative phase-contrast microscopy images showing NIH/3T3, PANC-1, and HT-1080 spheroids

reservoir, larger rectangular area on the bottom part is the sample chamber, filled with collagen gel seeded with randomly distributed cell spheroids. Gray scale along the image qualitatively represents chemoattractant diffusion from the reservoir to the collagen.

## 2.8 | Image and chemotaxis data analysis

The dynamic evolution of the tumor spheroid was quantified by measuring different morphological parameters[49,50] as shown in Figure 1C (right). Spheroids images were analyzed using a commercial image analysis software (Image Pro Plus 6.0, Media Cybernetics). Images were acquired every 20 min during TLM experiments, spheroids morphology was quantified every 12 h. The edge of the spheroid and the necrotic core (if present) were manually identified, measuring spheroid and necrotic core area. The thickness of viable rim ($\delta$, reported in figure) was calculated as the difference between spheroid and necrotic core equivalent radiuses. By approximating spheroids to ellipsoids, major and minor axis ($a$, and $b$ respectively, reported as dashed-dotted black lines), and orientation angle ($\theta°$) were analyzed. Two parameters were defined to relate the major and the minor axis: the deformation $\left(D = \frac{a-b}{a+b}\right)$ and the aspect ratio $\left(AR = \frac{a}{b}\right)$. Angles were measured as absolute value (no orientation) between the major axis and the vertical direction, which is parallel to the diffusive flow of chemoattractant. Spheroids are randomly orientated at the beginning of the experiment, in the range $0 < |\theta_0| < 90°$, where $\theta = 0°$ correspond to a major axis parallel to the gradient, as the y axis, while $\theta = 90°$ corresponds to spheroids oriented perpendicularly respect to the gradient direction, that is, parallel to the x axis. Evolution of $\theta$ as a function of time from its initial value can be due to gradient induced polarization of spheroids.

The same definition of angle was used to identify four different regions in each spheroid, depending to the exposure to the chemoattractant flow:

- Upwind: positive direction of y axis ±45°, identifying the fraction of spheroid edge facing the source of chemoattractant.
- Downwind: negative direction of y axis ±45°, corresponding to the 90° section of the spheroid edge opposite respect to the source of chemoattractant.
- Sides: positive or negative directions of x axis ±45°, corresponding to the two remaining lateral sections of the spheroid edge, facing the horizontal direction in our images. The chemoattractant diffusive flow is tangent to the edge in these two areas. Side sections are assumed to be exposed to identical chemical stimuli.

The three sections are indicated in white in Figure 1C and separated by white dashed lines. The cells eventually leaving each spheroid to invade the surrounding tissue were counted manually by direct visualization of the spheroid equatorial plane at different times (12, 18, 24, 36, and 48 h). In particular, cells of previous time were added to cells of specific time, in order to consider an integral quantity that may evaluate the whole response to the chemical stimulus. The number of invading cells was evaluated for each of the above defined three directions (upwind, downwind, sides). Cells were considered leaving the spheroid when they pass a threshold equal to 1.1 X the maximum radius measured of the spheroid over the entire experiment (spheroids tend to grow during the test).

To compare spheroids of different size, a linear cell flow was calculated by dividing the number of invading cells in each direction respect to the equatorial line of outflow corresponding to a 90° circular section (i.e., 0.5·$\pi$·R). The choice to calculate a linear flow is related to the fact the analysis is focused on the cells visualized on the spheroid equatorial plane, while cells out of focus originated at different latitudes of the spheroid, which might be visible by imaging the sample on different focus layers, are neglected. It is worth mentioning that results at least qualitatively comparable are obtained calculating surface flows (i.e., dividing the number of cells respect to an area of $1*\pi*R^2$ for each of the four directions), a part for higher error bars.

In isotropic conditions (ISO 0% and ISO 10%), the concentration driven stimulus does not depend on the position of the spheroid in the well, that is random. In the case of chemotactic condition ($\nabla C$), spheroids localized at different distances from the membrane can be subject a different chemotactic stimulus.[20] For this reason, in the analysis of invading cells, experimental data have been grouped into two categories, defined as low and high gradient (LG and HG, characterized by a specific gradient of $12 < SG < 22$ and $27 < SG < 40$ cm$^{-1}$), respectively considering the spheroid distance from the membrane.

Spheroid distance from the chamber side walls was also considered in the $\nabla C$ condition. In the case of spheroids too close to the chamber walls (i.e., below 0.9 mm, corresponding to about 15% of the chamber size along X direction), numerical simulations suggested the nutrient flow was not oriented along the Y direction only, due to side effects, inducing different stimuli in the two side directions. For this reason, spheroids with a distance from the walls lower than 0.9 mm were excluded from the analysis. For the same reason, spheroids that during the random deposition resulted too close each other were not considered in the analysis.

## 2.9 | Numerical analysis of nutrient diffusive flow

- Diffusive flow of glucose is considered here as typical representative of metabolites responsible for diffusional instability. In our previous works,[19,20] concentration profiles in the chemotaxis chamber are proved to be in agreement with a Fickian diffusion model. In this work, glucose concentration in the chemotactic chamber was analyzed by finite elements numerical simulation to calculate concentrations gradients in the sample well and estimate the chemotactic stimulus each spheroid was exposed to along the different directions (upwind, downwind, and side). The chemiotaxis chamber was discretized in 2D, following the scheme reported in Figure 1C. For each experiment, the size and position of each spheroid in the chemotactic chamber was determined by accurate visual inspection, measuring both the external diameter of the spheroid and that of the necrotic core (when present). Each spheroid was represented with two concentric circles, corresponding to the internal necrotic core and the external size, which also includes the viable rim, at initial time. The size of spheroids, and distribution among necrotic and viable areas, was assumed to remain unchanged during the entire simulation.

The diffusive transport of the chemoattractant in the chamber was described according to different parameters in each section of the domain, estimated from the literature for the system of our interest:

- in the chemoattractant reservoir: free diffusion of glucose in water ($D_g$=9.25·10-6 cm$^2$/s[51]);
- in the separation membrane: diffusion in a porous media (0.22 μm pores, tortuosity 2, according to data provided by the manufacturer; $D_g$=3.24·10$^{-6}$ cm$^2$/s);
- in the spheroid seeded sample reservoir: free diffusion of diluted glucose in collagen gel ($D_g$=7.63·10$^{-6}$ cm$^2$/s[48]);
- inside the spheroid, Fickian diffusion equation ($D_g$=1.24·10$^{-6}$ cm$^2$/s[51]) also includes glucose consumption in the viable rim, modeled according to a Michaelis–Menten kinetics[51] ($r = \frac{v_{max} \cdot c_G}{K_m + c_G}$, where $v_{max}$=0.01 mol/m$^3$s[51,52] is the maximum rate and $K_m$=8 mol/m$^3$ is the Michaelis–Menten constant.[52] This consumption of glucose was estimated referring to the initial size of the viable rim, as measured at $t = 0$.

As an example, chemoattractant concentration in the nearby of a spheroid is reported for the $\nabla C$ (Figure 2A,B, left) and ISO 10% condition (Figure 2A,B, right). Profile for ISO 0% condition is not reported here for brevity but is available in Supporting Information. As shown, in both cases, there is a drop in the concentration in correspondence

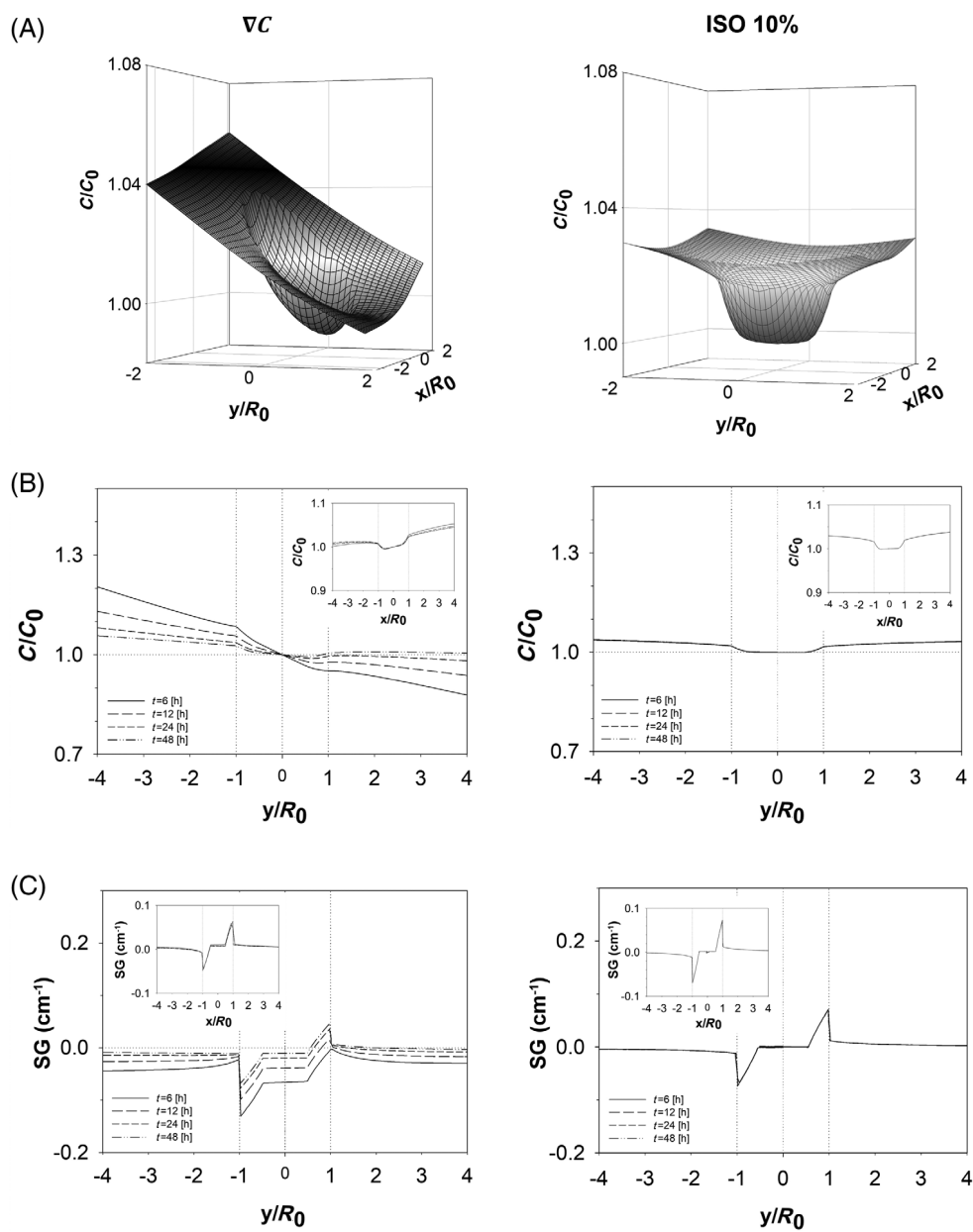

**FIGURE 2** Numerical simulation of nutrient concentration profile near the spheroid in $\nabla C$ (left) and ISO 10% (right) condition (profile for ISO 0% condition are analogous to normalized parameters reported for the case ISO 10%, and are not reported here for brevity, but are available in the Supporting Information). (A) 3D concentration profile at 24 h, the concentration drop due to the spheroid is qualitatively evident, together with a gradient related to the diffusive flow. (B) The nutrient concentration profile, normalized with respect to the concentration value at the center of the spheroid ($C_0$), along the gradient direction ($y/R_0$) is reported for different time values; in the inset, the same analysis is reported with respect to the x direction. (C) Specific gradient calculated considering the concentration value at the center of the spheroid, respect to the Y and X directions. Horizontal axes in B and C report the distance from the center of the spheroid, normalized with respect to the initial radius value of the spheroid $R_0$. Data are reported for three different times: 6, 24, 48 h. In all graphs, the dotted vertical lines represent the position of the spheroid

of the spheroid position, due to the consumption of nutrients. On top of this, in the ∇C case, there is an evident global trend in the concentration profile, that decreases with Y, and evolves in time, due to the transient bulk diffusion from the chemoattractant reservoir. This trend, in agreement with previous observations,[19,20] would be present also in the absence of spheroids, being described by the classical semi-infinite slab solution of Fick equation

$$\left( \frac{C}{C_0}(y,t) = \frac{1}{2}\left[1 - \text{erf}\left(\frac{y}{\sqrt{4Dt}}\right)\right] ; \; SG(y,t) \, (\text{cm}^{-1}) = \frac{dC(y,t)}{dy} \frac{1}{C(y,t)} = \frac{\exp\left(-\frac{y^2}{4Dt}\right)}{\sqrt{\pi Dt}\left[1 - \text{erf}\left(\frac{y}{\sqrt{4Dt}}\right)\right]} \right).$$

In our case, on top of this decreasing trend, there is also the consumption of nutrient that locally affects the profiles and the corresponding time evolution, resulting in the complex concentration and specific gradient profiles, reported in (Figure 2B,C) normalized respect to the concentration at the center of the spheroid. Data are plotted as function of the two main directions, normalized respect to the initial radius of the spheroid $R_0$. In the main chart of Figure 2B,C, data are reported as function of the Y direction, that is the direction of the bulk diffusive flow in the case of ∇C condition. In the inset of Figure 2B,C, the same quantities are reported as function of the X direction. Direct comparison of ∇C and ISO 10% conditions (left and right in Figure 2) shows that $C/C_0$ variation along X direction are negligible respect to the Y variation for ∇C condition (along a distance $\pm 2R_0$ the normalized concentration $C/C_0$ changes in the ranges [1–1.07] in the x direction, while the same displacement along y direction corresponds to a concentration range of [0.8–1.2], as visible from the chart). Similar result is observed about SG ($SG_x \in [-0.04$ to $0.06]$, while $SG_y \in [-0.13$ to $0.04]$). This proves the chemotactic stimulus on the spheroid is significantly different along the y direction respect to the X direction. For this reason, different morphology evolution of the spheroids under chemotactic stimuli is expected along the three directions above defined (upwind, downwind, sides). On the contrary, no significant difference between y and x direction is observed in the case of ISO 10%, the chemotactic stimulus being everywhere significantly lower compared to ∇C case. It is also worth mentioning negligible volume of spheroid respect to the entire chamber brings concentration profile to steady state on the time scale of observation in the ISO case, while time evolution is not negligible in the ∇C case, due to the intrinsically transient nature of the assay.

## 3 | STATISTICAL ANALYSIS

All experiments were performed at least two independent times with replicate samples for the ISO 10% condition. In each experiment, at least 15 spheroids were analyzed for each of the three conditions investigated (ISO 0%, ISO 10%, and ∇C). Data are expressed as mean ± standard error of mean (SEM). Comparisons between treated samples (ISO 0% and ∇C) and control (ISO 10%) were performed using one-tailed heteroscedastic Student's t-test. Results were considered statistically significant in the case of a $p$ value <0.05 (further detail of specific $p$ values are reported in figure captions).

## 4 | RESULTS

The assay here proposed provides an innovative tool that is the first systematic *in vitro* validation of the diffusional instability model. To stimulate cell metabolism and induce cancer cell invasion, glucose is used as chemoattractant. This experimental condition (∇C) was compared with two other conditions, the ISO 10%, to mimic standard cell growth, and ISO 0%, a nutrient-depleted environment. In the following section, the experimental results, obtained by applying our experimental method to different cell lines, are presented.

### 4.1 | NIH/3T3

Representative phase contrast microscopy images of NIH/3T3 spheroids are reported as a function of time in Figure 3. NIH/3T3 cells are fibroblasts from mouse embryos; being a non-tumor cell line, a limited

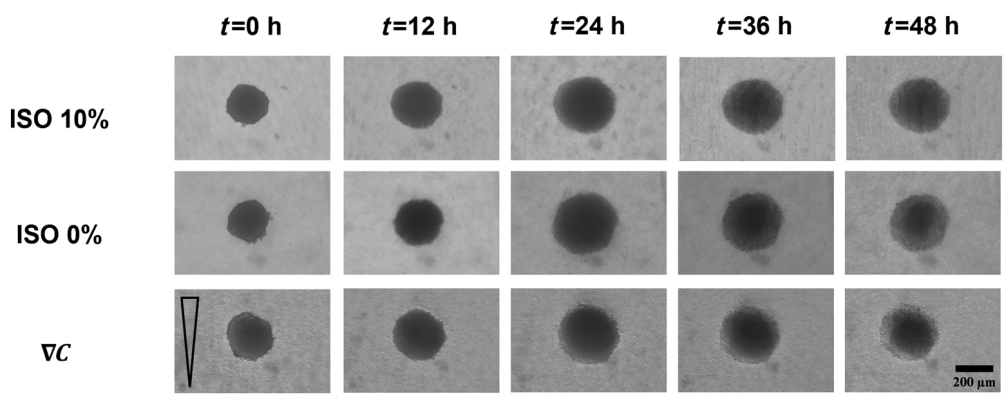

**FIGURE 3** Representative phase-contrast microscopy images showing the morphological response of NIH/3T3 spheroids at five different time points (0, 12, 24, 36, and 48 h) for three different conditions (ISO 10%, ISO 0%, and ∇C). Images in each row of the panels are relative to a single spheroid, taken as a qualitative representation of a given experimental condition. Scale bar: 200 μm

tendency to invade surrounding tissues is expected. ISO 10%, ISO 0%, and ∇C condition are compared along the rows, and each column of the figure panel is relative to a different time point (0, 12, 24, 36, and 48 h). In each row of the panel, a specific spheroid, representative of a given experimental condition, is shown at the time steps under investigation. Comparison of images at a given time point show qualitatively similar morphological response for each experimental condition.

In Figure 4A, the evolution of the spheroid area, normalized with respect to its initial value ($A/A_0$), is reported for each experimental condition. Spheroid area grows for the first 24 h with a similar trend under different conditions until it reaches an equilibrium condition. As can be seen, there is no significant difference between morphological responses in the absence or presence of a glucose chemoattractant stimulus: at 48 h, $A/A_0$ is about 2.09 ± 0.05 for ISO 10%, 1.83 ± 0.11 for ISO 0%, and 1.68 ± 0.07 for ∇C. A slightly higher growth is observed in ISO 10% condition, as expected under the standard growth condition, respect to the other two conditions where the spheroids are under nutrient stress. This result is an indication of a limited tendency of NIH/3T3 cell to invade the surrounding tissue, due to the non-tumor and non-invasive nature of this cell line.[53] This is attributed to the fact that individual cells did not leave the spheroids in any of the examined conditions due to the compact nature of the spheroid given significant cell–cell interactions.

NIH/3T3 spheroids showed a concentrically layered structure with a necrotic core surrounded by a viable rim of quiescent cells and an outer rim of proliferating cells, in agreement with the literature.[54,55] In Figure 4B, the size of the viable rim δ (i.e., the difference between the spheroid and necrotic core size) is reported for each experimental condition (ISO 10%, ISO 0%, and ∇C). Measurements show limited difference for isotropic and chemotaxis conditions (typical value of δ being about 45 μm, that can be assumed to be a measure of the diffusive distance of nutrients). In particular, ∇C conditions show a slightly smaller value in comparison to ISO conditions. This agrees with the differences observed in the size of the spheroid on the second day of the experiment and can be considered within the experimental noise.

In Figure 4C, the orientation angle is reported, and no significant differences among the three setups and no clear trends as function of time is observed in spheroid orientation with respect to the diffusive flow direction. The average value of θ at 48 h is about 50° for ISO 0% and ∇C conditions, and 60° for ISO 10% condition.

## 4.2 | PANC-1

In Figure 5, typical phase contrast microscopy images of PANC-1 spheroids are presented as a function of time, for ISO 10%, ISO 0%, and ∇C condition, reporting the same reference time points (0, 12, 24, 36, and 48 h). PANC-1 are human epithelial cells of carcinoma from pancreatic tissue. Being a tumor cell line, differences in comparison to the non-tumor case of NIH/3T3 are expected. Indeed, already a preliminary qualitative comparison of images shows different morphology of spheroids depending on the environmental conditions. Higher growth in the area in time for the ISO 0% and ∇C conditions, with respect to the ISO 10% condition is evident. Spheroids in ISO 10% subject to minimal diffusional stress are very compact, in comparison with the two other experimental conditions where the diffusional stress is heavier. In ISO 0% and ∇C, spheroids show an enlargement of the structure, losing compactness. Also in this case, individual cells did not leave the spheroid in none of the examined conditions.

The evolution of $A/A_0$ is reported for each experimental condition in Figure 6A. In this case, a significant difference is observed comparing the size evolution in the ISO 10%, where cells are in a nutrient reach environment, with respect to the ISO 0% and ∇C condition, where nutrient depletion causes diffusional instability of the tissue. Spheroid area grows for the first 24 h with a similar trend under different conditions. At longer times, in ISO 10%, the spheroid size reaches an equilibrium condition, with limited growth in time on the second day of the experiment. The size increment observed after 48 h is about 1.4×, 2.14×, and 2.36× for ISO 10%, ISO 0%, and ∇C conditions, respectively (with statistical difference of the stressed conditions with respect to the ISO 10% ($p < 5 \cdot 10^{-3}$ for ISO 0% and $p < 5 \cdot 10^{-4}$ for ∇C). This result can be attributed to a swelling of the spheroid structure, where tumor cells loose compactness and tend to escape from spheroid core, under stress, but without releasing cell–cell junctions, preserving the apparent morphology of an aggregate, probably because of a relevant cortical tension.

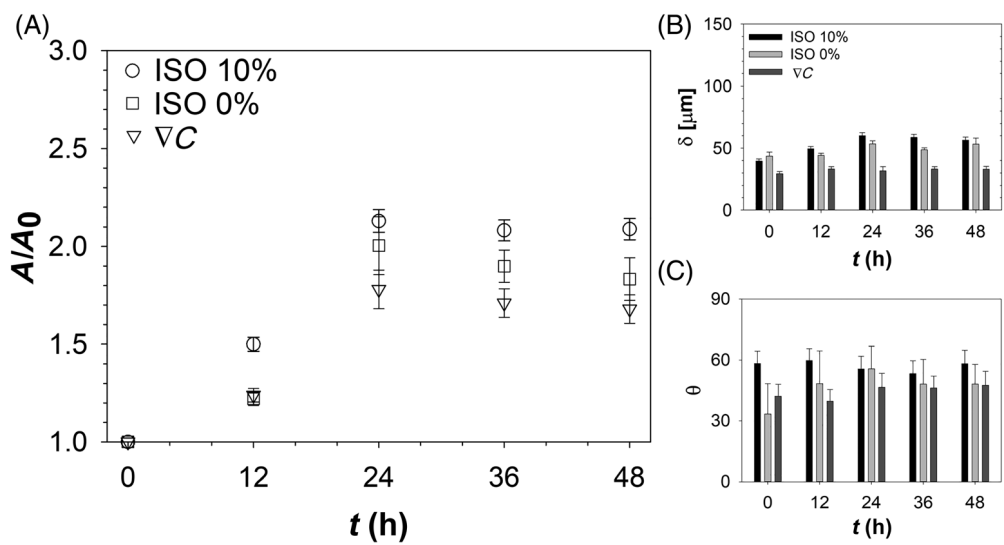

**FIGURE 4** (A) Evolution over time of the NIH/3T3 spheroid area, normalized with respect to its initial value ($A/A_0$), is reported for ISO 10%, ISO 0%, and ∇C conditions. (B) Evolution of the difference between the spheroid and necrotic core size, called "viable rim" δ, vs. time, is reported for each experimental condition. (C) The angle (θ) has been plotted as a function of time for each experimental condition

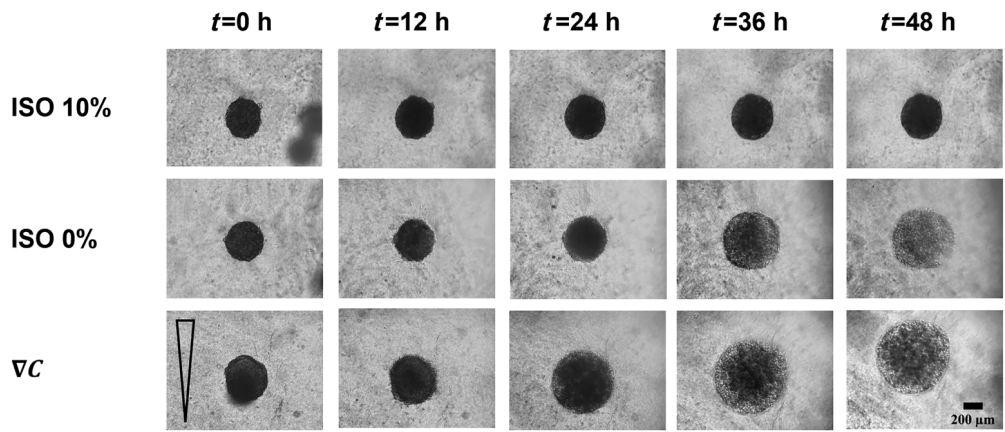

**FIGURE 5** Representative phase-contrast microscopy images showing the morphological response of PANC-1 spheroid at five different time points (0, 12, 24, 36, and 48 h) for three different conditions (ISO 10%, ISO 0%, and ∇C). Images in the panels are relative to a single spheroid, taken as qualitatively representative of a given experimental condition

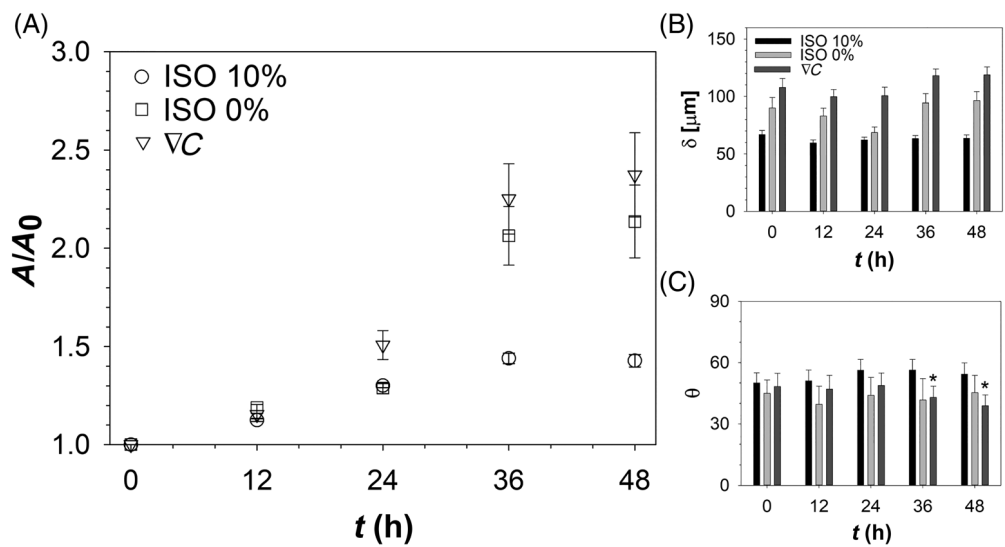

**FIGURE 6** (A) Evolution over time of the PANC-1 spheroid area, normalized with respect to its initial value ($A/A_0$), is reported for ISO 10%, ISO 0%, and ∇C conditions. (B) Evolution of the difference between the spheroid and necrotic core size, called viable rim δ, vs. time, is reported for each experimental condition. (C) The angle (θ) has been plotted as a function of time for each experimental condition (*$p < 0.05$)

In agreement with the NIH/3T3 case, also PANC-1 spheroids are characterized by the presence of a necrotic core, that allowed us to measure the size of the viable rim δ (Figure 6B). In the case of the ISO 10% condition, δ is about constant for the entire experiment, in agreement with the hypothesis of an equilibrium condition reached by the samples in the absence of diffusional stress. In stressed conditions (ISO 0% and ∇C), viable rim is significantly larger (1.22×), in agreement with a tendency of the living cells to expand to minimize the diffusional stress.

Angle does not show significant difference for the positive and negative control (ISO 10% and ISO 0%), with an average θ of about 45° (Figure 6C). In the case of ∇C conditions, significant differences are observed ($p < 0.05$, calculated with respect to ISO 10%), with a reduction in the measured average value at long times, corresponding to a preferential orientation of spheroid major axis toward the direction of the chemoattractant gradient.

### 4.3 | HT-1080

Typical phase contrast microscopy images of the spheroid as a function of time (0, 12, 24, 36, and 48 h) are presented in Figure 7 for HT-1080, where samples in ISO 10%, ISO 0%, and ∇C conditions are compared. Images show qualitatively a significant difference in the sample morphology as a function of the experimental conditions. A relevant number of cells leave the spheroids invading the surrounding matrix, and an elongation of the spheroid along the gradient direction is visible for the ∇C condition. Time-lapse videos of morphological response of HT-1080 cell spheroids to three different conditions are reported in Videos S2–S4.

The evolution of $A/A_0$ is reported for each experimental condition in Figure 8A. For ISO 0% and ∇C, spheroid area grows for the first 24 h until it reaches an equilibrium condition, in agreement with the other two cell lines analyzed; on the contrary, for ISO 10%, the area value reaches an equilibrium condition (~1.10) after only 12 hours. As shown, $A/A_0$ value is higher for ISO 0% and $\nabla C$ compared to the ISO 10% condition, in particular, 1.5× and 1.3×, respectively ($p < 0.05$ for ISO 0% and $p < 5 \cdot 10^{-5}$ for $\nabla C$). This result is qualitatively in agreement with results of PANC-1 samples, where a swelling in the cell aggregate was induced by the diffusional stress. Moreover, an elongation of spheroid shape in the direction of the gradient, in the $\nabla C$ condition, is qualitatively evident in Figure 7. To quantify this elongation, the deformation (D) and aspect ratio (AR) approximating the spheroid area with an ellipsoid were calculated. As shown in Figure 8B, deformation parameter normalized respect to its initial value ($D/D_0$), is significantly ($p < 0.05$) higher for the $\nabla C$ condition compared to the ISO 10% condition (1.8× at 48 h). As shown in the Figure 8C, the AR also increases over time for the $\nabla C$ condition: the major axis, a, being about 10% higher than the minor axis, b, at 48 h. Also, this parameter is almost constant over time for the ISO 10% condition. As shown in Figure 8D, the value of $\theta$ for the $\nabla C$ condition is significantly lower with respect to isotropic conditions (20° for $\nabla C$ condition and about 60° for the two isotropic conditions). Having defined the spheroids orientation as the angle between the major axis of spheroid, a, and the y axis, that is, the direction of the gradient, the reduction of $\theta$ over time for $\nabla C$ condition confirms the polarization of the spheroid in the direction of the chemotactic gradient induced by diffusional instability mechanisms. Moreover, due to the gradient, a cell flow leaving the spheroid was observed. A detailed quantitative analysis was performed by measuring the number of cells leaving the spheroids in each direction. In the case of limited diffusional stress (ISO 10%), almost no cells (less than four cells for each spheroid) were observed to leave the spheroids during the entire experiment (data not shown). On the contrary, in the *negative control*, ISO 0% condition, a significant number of cells was observed to leave spheroids during the experiment. On average, 124 ± 86 cells were measured leaving each spheroid along its diametral plane after 48 h. In the case of $\nabla C$ condition, spheroids were grouped in two categories defined as LG and HG, with respect to the intensity of the chemotactic stimulus imposed (see above). As a generalized result, on average, after 48 h 397 ± 111 cells left each spheroid subjected to LG and 425 ± 46 cells for HG conditions ($p < 0.05$), independently from spheroid size and direction of migration. Cell flow for the ISO 0% condition is ~3× lower respect to $\nabla C$ condition. It is worth mentioning this difference could appear in contrast with the similarity in the two concentration profiles reported in Figure 2 after 48 h. It should be considered that cells flow measured at a given time (e.g., 48 h) is the result of the entire history of chemical stimulus cells were subject to from the beginning of the experiment up to that time. More detailed analysis of this aspect is reported in Supporting Information.

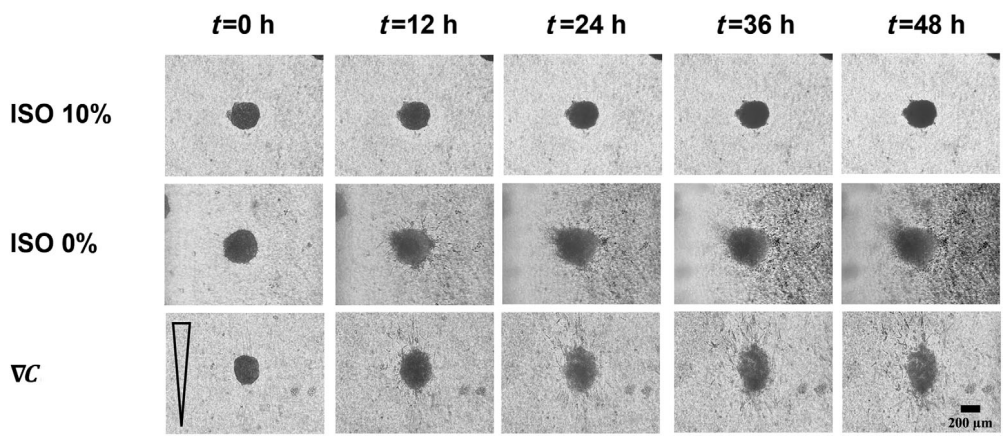

**FIGURE 7** Representative phase-contrast microscopy images showing the morphological response of HT-1080 spheroid at five different time points (0, 12, 24, 36, and 48 h) for three different conditions (ISO 10%, ISO 0%, and $\nabla C$). Images in the panels are relative to a single spheroid, taken as qualitatively representative of a given experimental condition. Representative time-lapse videos–related HT-1080 cell spheroids for ISO 10%, ISO 0%, and $\nabla C$ are included in Videos S2–S4

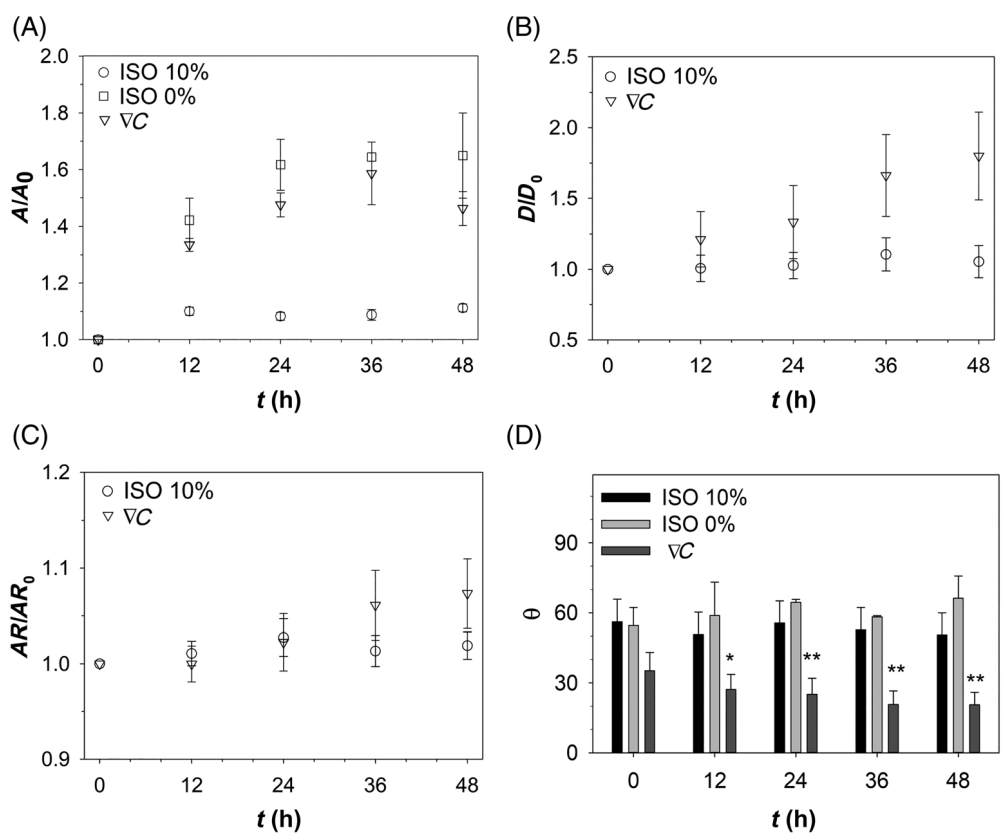

**FIGURE 8** Evolution over time of the HT-1080 spheroid area (A), deformation (B), and aspect ratio (C) normalized with respect to its initial value) is reported for each experimental condition. (D) Evolution of the angle ($\theta$) as a function of time for each experimental condition (*$p < 0.05$, **$p < 0.01$)

In order to quantify the presence of any preferential direction in the invasion of the surrounding tissue, cells leaving the spheroids along each of the three directions (upwind, downwind, side) were identified. To compare spheroids of different size, a linear cellular flow (defined above) was also calculated.

In Figure 9A, the flow of cells coming out through the upwind, downwind, and side is reported as function of time for the ISO 0% case. Data show that the flows along the direction are comparable (about 0.09 cells/μm h), with limited growth in time. No preferential direction is observed, in agreement with the isotropic condition of the *negative control* setup.

In Figure 9B,C, the cell flow along the three directions is reported for the LG and HG conditions, respectively. Data show that, for the $\nabla C$ condition, the cell flow along the upwind direction (on average $1.30 \cdot 10^3$ and $1.73 \cdot 10^3$ cells/μm h for LG and HG) is only slightly higher with respect to the downwind direction ($1.26 \cdot 10^3$ and $1.58 \cdot 10^3$ cells/μm h), in agreement with the predictions from the numerical simulation that indicate a comparable value of the specific gradient in these two regions of the spheroid (SG = 3.9 and $2.5 \cdot 10^{-2}$ cm$^{-1}$, in absolute value, for upwind and downwind, respectively, after 6 h, and SG = 1.0 and $1.3 \cdot 10^{-3}$cm$^{-1}$ after 48 h, measured as example of typical values at one radius from the spheroid edge, data extracted from Figure 2). On the other hand, the cell flow along the two side directions, is significantly lower (on average 684 and 828 cells/μm h for the LG and HG conditions, respectively). This agrees with the estimates of the lower gradient along X direction, as reported in Figure 2 (SG = 1.1 and $9.1 \cdot 10^{-3}$cm$^{-1}$ after 6 h while SG = 8.3 and $0.75 \cdot 10^{-4}$cm$^{-1}$ after 48 h for the two sides, respectively, that are here assumed to be equivalent).

Cells flow in $\nabla C$ condition as response to chemical stimulus is fundamental to describe the AR profile, in particular at 24 h. In fact, cells leaving the spheroids are significantly higher in the upwind and downwind directions (corresponding to Y) respect to side (X direction), even in the initial steps of the experiment. The cell flow induces a non-isotropic erosion of the spheroids that balances the elongation along the main direction. For this reason, in the initial steps, the differences in AR among $\nabla C$ and both ISO conditions are not significant, while appear to be measurable at longer times. This behavior is aided by the fibroblast-like nature of HT-1080 cells, which tend to migrate into the matrix as individual cells, resulting in a stronger invasiveness, respect to the case of PANC-1, where tight intracellular junctions[56] limit the tendency of individual cells to invade the ECM.

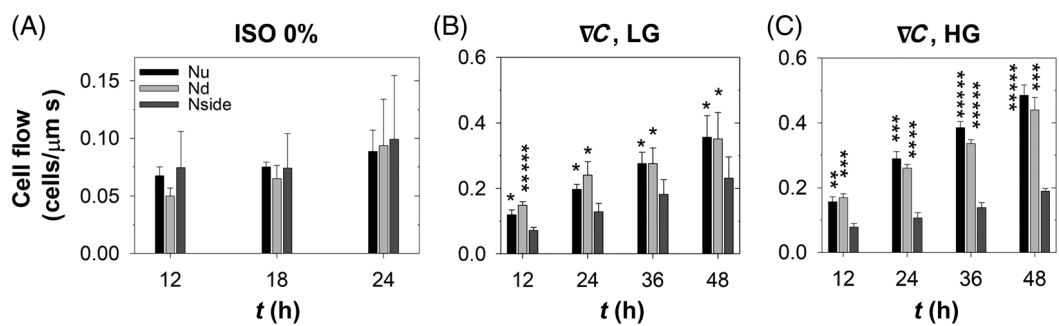

**FIGURE 9** Cells flow leaving the HT-1080 spheroid for ISO 0% and $\nabla C$ (low and high gradient conditions) in each direction (upwind, downwind, and side). No significant differences are observed in ISO 0%. In the two $\nabla C$ conditions (diagrams B and C), cell flow in side direction ($N_{side}$) is compared to flow in upwind and downwind ($N_u$ and $N_d$). Statistical significance of the difference among $N_u$ respect to $N_{side}$ and $N_d$ respect to $N_{side}$, calculated as p value, is reported as "*" superimposed on the bars (*$p < 0.05$, **$p < 0.05$, ***$p < 5 \cdot 10^{-3}$, ****$p < 1 \cdot 10^{-3}$, *****$p < 5 \cdot 10^{-4}$)

## 5 | SUMMARY AND DISCUSSION

In this paper, an innovative experimental investigation aimed to quantify the role of transport phenomena on cancer invasion is proposed. The methodology is based on the diffusional instability model[1,15] and can be used to study systematically *in vitro* and *ex vivo* 3D tumor models.

*In vitro* dynamic evolution of cancer spheroids subjected to an external chemical stimuli[19,20] is here investigated. Tumor spheroids were suspended in a 3D collagen matrix and observed for 48 h by means of time-lapse video microscopy. In fact, by using a chemotaxis chamber,[19,20] it was possible to impose a controlled diffusive gradient of glucose to the spheroids, mimicking the physiological conditions present *in vivo* where an avascular tumor aggregate is exposed to nutrient flows coming from nearby vessels.

To compare the response of different invading cancers to the chemotactic stimulus, different immortalized cell lines were used, intended to be representative of the most common types of cancer: PANC-1 and HT-1080. For cell lines here investigated, the chemotactic response to a steady spatial concentration gradient is reported only for single cells, for different chemoattractants,[57–63] not for a 3D model. The NIH/3T3 non-tumor cell line is considered as reference control system. As expected, NIH/3T3 spheroids preserve a spheroidal shape and cells do not leave the spheroid. This result is a consequence of the limited tendency of NIH/3T3 cell to invade the surrounding tissue, due to the non-tumor and non-invasive type of this cell line. On the other hand, spheroids of HT-1080 cancer cell lines, according to their fibroblast-like nature, appeared to be characterized by a high-invasive phenotype: a relevant number of cells left spheroids and an elongation of cell spheroid along the gradient direction (Y axis) was visible for the chemotaxis condition. In other terms, the chemical energy,[38] due to the glucose gradient, allows to overcome the cortical tension of cell spheroid, making cells free to migrate and invade the surrounding tissue breaking the symmetry of the system. Moreover, HT-1080 spheroids showed a limited increase in area, that is also a direct consequence of the mass loss due to cell motility, which was not adequately balanced by an increase in the proliferation.[64] In other terms, in our experimental setup, chemical stimuli significantly influenced cell motility, but only marginally cell proliferation. On the contrary, PANC-1 spheroids showed a lower invasive behavior: higher growth in area is evident, but individual cells did not leave spheroid in none of examined conditions. Probably, this is due to a limited chemotactic stimulus and a low concentration of glucose,[65] that was used as chemoattractant. In fact, it is known that high glucose concentration, after about 48 h, is expected to induce epithelial-mesenchymal transition (EMT) in epithelial cell lines, by which cells lose epithelial cell–cell adhesion with improved tendency to move through extracellular matrix.[66] In the transition from epithelial to mesenchymal phenotype, cells not only change morphology but increase their motility, modulating tumor cell invasion and metastasis.[67–70] EMT and consequently increased motility was not observed in our experiments, probably due to the insufficient concentration of glucose. A further, more detailed, investigation of this aspect could be done by marking adhesion molecules using classical biochemical techniques.[70]

The approach here proposed represents an innovative technique to investigate complex biological systems, such as cancer. The possibility to quantify differences in invasiveness of cancer cell types and the analysis of the role of transport phenomena, typical of a chemical engineering approach, represents a key advancement that can be implemented in personalized medicine applications. In fact, the *in vitro* assay here proposed and validated in the case of tumor spheroids of immortalized cells can be applied to monitor diffusional instability of cancer organoids[71] developed from primary cells, also obtained from patients. This *ex vivo* application can allow to reproduce in the laboratory response of patient specific cultures to different pharmacological treatments, directly impact on drug efficacy and safety studying the drug–receptor binding kinetics,[72] in order to design precision treatments.

### ACKNOWLEDGMENTS

Speranza Esposito and Valeria Rachela Villella are gratefully acknowledged for their valuable support to the cell cultures. Brunella Cipolletta, Valerio Darino, Marica De Luca, Martina Schibeci, and Elena Scuncio contributed to the experiments and data analysis during their master and bachelor theses. PD acknowledges the Cockrell Foundation for their financial support. This study was conducted under the umbrella of the International Academic Affiliation Agreement between Houston Methodist Academic Institute (Houston, TX, USA) and University of Naples Federico II (Napoli, Italy). Open Access Funding provided by Universita degli Studi di Napoli Federico II within the CRUI-CARE Agreement.

### AUTHOR CONTRIBUTIONS

**Rosalia Ferraro:** Data curation (equal); formal analysis (equal); investigation (equal); software (equal); writing – original draft (lead). **Flora Ascione:** Formal analysis (equal); investigation (equal); methodology (equal); software (equal); visualization (equal). **Prashant Dogra:** Validation (equal); visualization (equal). **Vittorio Cristini:** Conceptualization (equal); validation (equal). **Stefano Guido:** Conceptualization (equal); resources (lead); validation (equal). **Sergio Caserta:** Conceptualization (equal); data curation (equal); investigation (equal); methodology (lead); software (lead); supervision (lead); validation (equal); writing – review and editing (lead).

### DATA AVAILABILITY STATEMENT

The data that support the findings of this study are available from the corresponding author upon reasonable request.

### ORCID

*Sergio Caserta*  https://orcid.org/0000-0002-4400-0059

## SUPPORTING INFORMATION

Additional supporting information may be found in the online version of the article at the publisher's website.